\documentclass{article}
\usepackage{latexsym}

\newcommand{\newc}{\newcommand}

\newc{\beq}{\begin{equation}}
\newc{\eeq}{\end{equation}}
\newc{\beqarray}{\begin{eqnarray}}
\newc{\eeqarray}{\end{eqnarray}}
\newc{\eq}[1]{Eq.~(\ref{#1})}
\newc{\order}[1]{{{\cal O}(#1)}}
\newc{\Ref}[1]{Ref.~\cite{#1}}
\newc{\fig}[1]{Fig.~\ref{#1}}
\newc{\vev}[1]{\left\langle {#1}\right\rangle}
\newc{\ket}[1]{\left| #1 \right\rangle}
\newc{\bra}[1]{\left\langle #1 \right|}
\newc{\matel}[3]{\left\langle #1 | #2 | #3 \right\rangle}
\newc{\QQbar}{Q\overline{Q}}
\newc{\Qqbar}{Q\overline{q}}
\newc{\qQbar}{q\overline{Q}}
\newc{\qqbar}{q\overline{q}}
\newc{\phiQQ}{\phi_{\scriptscriptstyle\rm\QQbar}}
\newc{\VQQ}{V_{\scriptscriptstyle\rm\QQbar}}
\newc{\gauge}{A^+ = 0}
\newc{\gN}{g_{\scriptscriptstyle\rm N}}
\newc{\gNsq}{g^2_{\scriptscriptstyle\rm N}}
\newc{\mQ}{m_{\scriptscriptstyle\rm Q}}
\newc{\mq}{m_{\scriptscriptstyle\rm q}}
\newc{\MQq}{M_{\scriptscriptstyle\rm Qq}}

\begin{document}

\title{
\hspace*{8cm} {\large SMUHEP 99-11}\\
\hspace*{7.7cm} {\large hep-ph/9910312}\\
\vspace*{.8cm}
String Breaking in Two-Dimensional QCD\footnote{
 Contribution to the Proceedings of the TJNAF Workshop on the 
 Transition from Low to High Q Form Factors, in honor of
 Stanley Brodsky's 60th birthday, Athens, GA, September 17, 1999.}
}

\author{K.\ Hornbostel\\[.2cm]
\small Southern Methodist University, Dallas, TX 75275
}

\date{September 1999}

\maketitle

\begin{abstract}
I present results of a numerical calculation of the effects of light 
quark-antiquark pairs on the linear heavy-quark potential in light-cone
quantized two-dimensional QCD.  I extract the potential from the
$\QQbar$ component of the ground-state wavefunction, and observe 
string breaking at the heavy-light meson pair threshold.  I briefly
comment on the states responsible for the breaking.
\end{abstract}

\vfill

\pagebreak

\section{Introduction}

Pure Quantum Chromodynamics exhibits linear confinement between
charges at large distances, binding them with a gluonic string,
as observed some time ago in lattice simulations of the
potential between static quarks \cite{creutz}.  Stretching the string 
sufficiently in the presence of light quarks should produce quark-antiquark 
pairs, breaking the string and screening the charges; attempts to clearly 
demonstrate this effect in lattice QCD are ongoing \cite{schilling}.  
The purpose for this study is to determine 
whether string breaking can be observed in a relatively simple model, 
two-dimensional QCD \cite{thomod}, using a very different numerical scheme: 
solving for eigenstates of the light-cone (LC) Hamiltonian for the system 
confined to a box \cite{thesis}.

Early interest in LC quantization focused on high-energy
scattering, due in part to the boost invariance of LC wavefunctions 
and their direct connection to parton distribution functions 
\cite{exclproc}.  However, the suggestion of a simplified
vacuum and successes with two-dimensional models generated some 
recent interest in using this scheme to perform nonperturbative 
calculations of static properties in QCD \cite{paubrod}.  In addition to 
their connection to parton distributions, LC wavefunctions have a 
well-defined heavy-quark limit \cite{burkHVQ} and a simple connection 
to quantum mechanical wavefunctions, which I will make use of in 
extracting the heavy-quark potential below.

In this formalism, one initializes fields and their commutators at
equal LC time, $x^+ \equiv x^0 + x^3 = 0$.  The Hamiltonian 
$P^- \equiv P^0 - P^3$ evolves states in $x^+$, while the LC momentum 
$P^+ \equiv  P^0 + P^3$ is kinematic and conserved in interactions. 
Diagonalizing $P^-$ in the space of states defined by acting with
creation operators on the vacuum produces the masses and wavefunctions
for hadrons and multiple-hadron states.  All hadronic matrix elements
may be expressed as integrals over these wavefunctions.

In order to accomplish this numerically, I confine the system to
a box of length $L$ in $x^-$ and impose antiperiodic boundary conditions
on the fields.  As a result, each quark carries $k^+$ in odd
half integer units $2\pi/L$.  Because all $k^+$ are positive and 
must sum to the total $P^+$, the discrete momentum $K \equiv (L/2\pi) P^+$
determines both the resolution of the wavefunctions and also controls
the number of quanta possible in a state.  $K \rightarrow \infty$ defines
the continuum limit.

The program which simulates this system allows for any number of colors 
and flavors.  Once given conserved quantities such as baryon number, 
total momentum, and individual flavor numbers, it generates all states 
consistent with these, though it allows for additional restrictions.  
It computes the Hamiltonian in this basis and diagonalizes it, generating 
the complete set of masses and 
wavefunctions.  That it produces all states is both an advantage and curse, 
as the number of states grows roughly exponentially with $K$.  See J.~Hiller's
contribution to these proceedings \cite{hiller} and references therein for 
a discussion of an intelligent way to focus on the lowest lying states.  

While fully relativistic, the use of a Hamiltonian, a Fock space description
of states, and wavefunctions makes it similar in form to nonrelativistic
many-particle quantum mechanics.  A final similarity is the ability to 
work with a simple vacuum.  Roughly, because all particles carry positive
$k^+$, it is not possible for their momentum to sum to the zero $P^+$ of the
vacuum; therefore, the vacuum has no particles.  The true situation is
more involved, and whether this argument holds depends on what is being 
computed.  Extracting vacuum properties in particular can be subtle \cite{vac}.
But when applicable it is an enormous simplification.

One of the advantages of LC quantization absent in the
older infinite momentum frame formulation is that it selects
a quantization surface rather than a specific reference frame.
As a result, it has a well-defined nonrelativistic limit
for heavy quarks, in which LC wavefunctions go smoothly 
into the usual wavefunctions of nonrelativistic quantum mechanics.
Because I wish to extract the potential felt by two nearly
static quarks, I include only one heavy flavor.  Heavy in
two-dimensional QCD means that $\mQ$ is large compared to the 
dimensionful coupling $g$; in four dimensions, the comparison would 
be to $\Lambda_{\rm QCD}$.

To see the Schr\"odinger equation these satisfy, consider the 
projection of the LC equation 
\beq
M^2\ket{\phi} = P^+ P^-\ket{\phi} 
\eeq
onto the $\ket{\QQbar}$ subspace.  States in this subspace have
the form
\beq
\int_0^1 dx\, \phiQQ(x)\, b^{c\dagger}(x P^+)\, d^\dagger_c((1-x)P^+)\ket{0}
\eeq
where $x$ gives the fraction of $P^+$ carried by the quark
with probability $|\phiQQ(x)|^2$, and the color index $c$ contracts
to form a singlet.  While the complete wavefunction 
contains states with extra $\QQbar$ pairs, these are suppressed by 
factors of $1/\mQ$.
The result is 't Hooft's equation \cite{thomod}:
\beq
\label{eq:tHooft}
 M^2\phiQQ(x) = \mQ^2\left[{1\over x} + {1\over (1-x)}\right]\phiQQ(x)
            - {\gNsq\over\pi} \int_0^1 dy {\phiQQ(y) - \phiQQ(x)\over (y-x)^2} \; .
\eeq
Here $\gNsq \equiv g^2 (N^2-1)/2N$, and the principal value prescription
defines the integral.  Though restricted to the two quark
sector, this equation is still fully relativistic; in fact, it describes
mesons to leading order in large $N$.

In the large-$\mQ$ limit, the momentum fraction carried by the quark,
\beq
x \equiv {k^+ \over P^+} = {E(k) + k\over E(P) + P} \approx {1\over 2} + {q\over 2\mQ}
\eeq
to lowest order in the relative momentum $q$, and the LC
wavefunction $\phiQQ(x)$ becomes sharply peaked around $1/2$.  
Introducing a nonrelativistic momentum-space wavefunction
\beq
\label{eq:psi}
 \psi(q) \equiv \phiQQ(1/2 + q/2\mQ)
\eeq
focuses on deviations from this peak.  Defining the nonrelativistic 
energy $E \equiv M - 2\mQ$ and reduced mass $\mu \equiv \mQ/2$,
extending $q$ from $\pm\mQ$ to infinity, and Fourier transforming
$\psi(q)$ to the relative position $r$ turns \eq{eq:tHooft} into 
the Schr\"odinger equation \cite{hamer, burkqcd2, thesis}
\beq
\label{eq:schr}
\left[-{1\over 2\mu}\partial_r^2 + 
   {1\over 2} \gNsq |r| \right]\psi(r) = E\psi(r) \; .
\eeq
The relation
\beq
 \int_\infty^\infty {dq\over q^2} [e^{-i q r} - 1] = -\pi |r| \;.
\eeq
converted the integral over $q$ into a linear potential in position 
with string tension $\gNsq/2$.  The solutions are Airy functions; see
\Ref{hamer} for details.

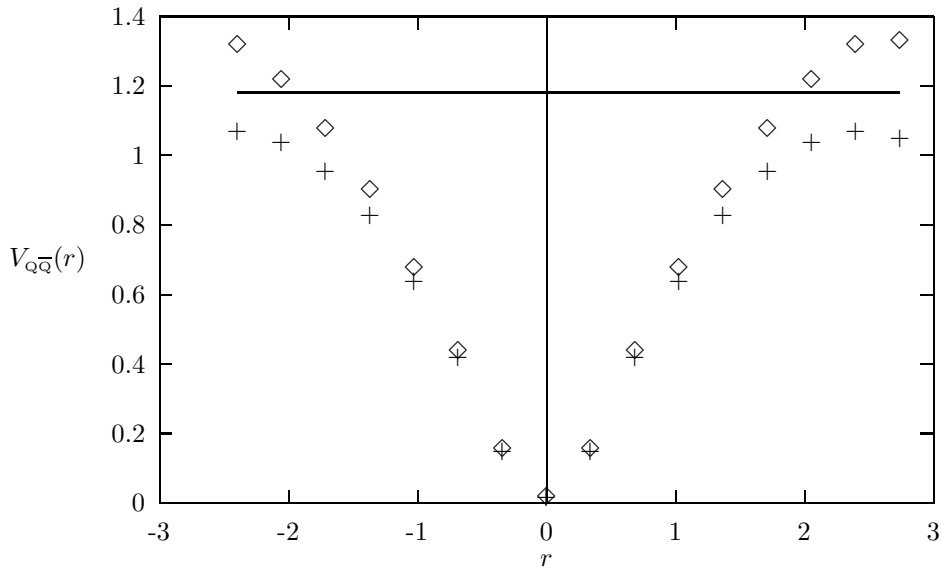
\begin{figure}
\setlength{\unitlength}{0.240900pt}
\ifx\plotpoint\undefined\newsavebox{\plotpoint}\fi
\sbox{\plotpoint}{\rule[-0.200pt]{0.400pt}{0.400pt}}%
\begin{picture}(1500,900)(0,0)
\font\gnuplot=cmr10 at 10pt
\gnuplot
\sbox{\plotpoint}{\rule[-0.200pt]{0.400pt}{0.400pt}}%
\put(220.0,113.0){\rule[-0.200pt]{292.934pt}{0.400pt}}
\put(828.0,113.0){\rule[-0.200pt]{0.400pt}{184.048pt}}
\put(220.0,113.0){\rule[-0.200pt]{4.818pt}{0.400pt}}
\put(198,113){\makebox(0,0)[r]{0}}
\put(1416.0,113.0){\rule[-0.200pt]{4.818pt}{0.400pt}}
\put(220.0,222.0){\rule[-0.200pt]{4.818pt}{0.400pt}}
\put(198,222){\makebox(0,0)[r]{0.2}}
\put(1416.0,222.0){\rule[-0.200pt]{4.818pt}{0.400pt}}
\put(220.0,331.0){\rule[-0.200pt]{4.818pt}{0.400pt}}
\put(198,331){\makebox(0,0)[r]{0.4}}
\put(1416.0,331.0){\rule[-0.200pt]{4.818pt}{0.400pt}}
\put(220.0,440.0){\rule[-0.200pt]{4.818pt}{0.400pt}}
\put(198,440){\makebox(0,0)[r]{0.6}}
\put(1416.0,440.0){\rule[-0.200pt]{4.818pt}{0.400pt}}
\put(220.0,550.0){\rule[-0.200pt]{4.818pt}{0.400pt}}
\put(198,550){\makebox(0,0)[r]{0.8}}
\put(1416.0,550.0){\rule[-0.200pt]{4.818pt}{0.400pt}}
\put(220.0,659.0){\rule[-0.200pt]{4.818pt}{0.400pt}}
\put(198,659){\makebox(0,0)[r]{1}}
\put(1416.0,659.0){\rule[-0.200pt]{4.818pt}{0.400pt}}
\put(220.0,768.0){\rule[-0.200pt]{4.818pt}{0.400pt}}
\put(198,768){\makebox(0,0)[r]{1.2}}
\put(1416.0,768.0){\rule[-0.200pt]{4.818pt}{0.400pt}}
\put(220.0,877.0){\rule[-0.200pt]{4.818pt}{0.400pt}}
\put(198,877){\makebox(0,0)[r]{1.4}}
\put(1416.0,877.0){\rule[-0.200pt]{4.818pt}{0.400pt}}
\put(220.0,113.0){\rule[-0.200pt]{0.400pt}{4.818pt}}
\put(220,68){\makebox(0,0){-3}}
\put(220.0,857.0){\rule[-0.200pt]{0.400pt}{4.818pt}}
\put(423.0,113.0){\rule[-0.200pt]{0.400pt}{4.818pt}}
\put(423,68){\makebox(0,0){-2}}
\put(423.0,857.0){\rule[-0.200pt]{0.400pt}{4.818pt}}
\put(625.0,113.0){\rule[-0.200pt]{0.400pt}{4.818pt}}
\put(625,68){\makebox(0,0){-1}}
\put(625.0,857.0){\rule[-0.200pt]{0.400pt}{4.818pt}}
\put(828.0,113.0){\rule[-0.200pt]{0.400pt}{4.818pt}}
\put(828,68){\makebox(0,0){0}}
\put(828.0,857.0){\rule[-0.200pt]{0.400pt}{4.818pt}}
\put(1031.0,113.0){\rule[-0.200pt]{0.400pt}{4.818pt}}
\put(1031,68){\makebox(0,0){1}}
\put(1031.0,857.0){\rule[-0.200pt]{0.400pt}{4.818pt}}
\put(1233.0,113.0){\rule[-0.200pt]{0.400pt}{4.818pt}}
\put(1233,68){\makebox(0,0){2}}
\put(1233.0,857.0){\rule[-0.200pt]{0.400pt}{4.818pt}}
\put(1436.0,113.0){\rule[-0.200pt]{0.400pt}{4.818pt}}
\put(1436,68){\makebox(0,0){3}}
\put(1436.0,857.0){\rule[-0.200pt]{0.400pt}{4.818pt}}
\put(220.0,113.0){\rule[-0.200pt]{292.934pt}{0.400pt}}
\put(1436.0,113.0){\rule[-0.200pt]{0.400pt}{184.048pt}}
\put(220.0,877.0){\rule[-0.200pt]{292.934pt}{0.400pt}}
\put(45,495){\makebox(0,0){$V_{\scriptscriptstyle\rm Q\overline{Q}}(r)$}}
\put(828,23){\makebox(0,0){$r$}}
\put(220.0,113.0){\rule[-0.200pt]{0.400pt}{184.048pt}}
\put(343,833){\raisebox{-.8pt}{\makebox(0,0){$\Diamond$}}}
\put(412,778){\raisebox{-.8pt}{\makebox(0,0){$\Diamond$}}}
\put(481,701){\raisebox{-.8pt}{\makebox(0,0){$\Diamond$}}}
\put(551,605){\raisebox{-.8pt}{\makebox(0,0){$\Diamond$}}}
\put(620,482){\raisebox{-.8pt}{\makebox(0,0){$\Diamond$}}}
\put(689,351){\raisebox{-.8pt}{\makebox(0,0){$\Diamond$}}}
\put(759,197){\raisebox{-.8pt}{\makebox(0,0){$\Diamond$}}}
\put(828,122){\raisebox{-.8pt}{\makebox(0,0){$\Diamond$}}}
\put(897,197){\raisebox{-.8pt}{\makebox(0,0){$\Diamond$}}}
\put(967,351){\raisebox{-.8pt}{\makebox(0,0){$\Diamond$}}}
\put(1036,482){\raisebox{-.8pt}{\makebox(0,0){$\Diamond$}}}
\put(1105,605){\raisebox{-.8pt}{\makebox(0,0){$\Diamond$}}}
\put(1175,701){\raisebox{-.8pt}{\makebox(0,0){$\Diamond$}}}
\put(1244,778){\raisebox{-.8pt}{\makebox(0,0){$\Diamond$}}}
\put(1313,833){\raisebox{-.8pt}{\makebox(0,0){$\Diamond$}}}
\put(1383,839){\raisebox{-.8pt}{\makebox(0,0){$\Diamond$}}}
\put(343,696){\makebox(0,0){$+$}}
\put(412,679){\makebox(0,0){$+$}}
\put(481,634){\makebox(0,0){$+$}}
\put(551,565){\makebox(0,0){$+$}}
\put(620,461){\makebox(0,0){$+$}}
\put(689,342){\makebox(0,0){$+$}}
\put(759,194){\makebox(0,0){$+$}}
\put(828,122){\makebox(0,0){$+$}}
\put(897,194){\makebox(0,0){$+$}}
\put(967,342){\makebox(0,0){$+$}}
\put(1036,461){\makebox(0,0){$+$}}
\put(1105,565){\makebox(0,0){$+$}}
\put(1175,634){\makebox(0,0){$+$}}
\put(1244,679){\makebox(0,0){$+$}}
\put(1313,696){\makebox(0,0){$+$}}
\put(1383,686){\makebox(0,0){$+$}}
\sbox{\plotpoint}{\rule[-0.400pt]{0.800pt}{0.800pt}}%
\put(343,757){\usebox{\plotpoint}}
\put(343.0,757.0){\rule[-0.400pt]{250.536pt}{0.800pt}}
\end{picture}
\caption{
 The potential defined in \eq{eq:VQQ} at $\gN/\mQ = .164$ for two heavy 
 quarks ($\Diamond$), and after the inclusion of one light flavor with
 $\mq/\mQ = .001$ ($+$).  The discrete momentum $K=15$ for both cases.  
 The horizontal line gives the additional energy $2\MQq - 2\mQ$ required 
 to produce two heavy-light mesons, each computed at $K=8$.  The square
 root of the string tension $\gNsq/2$ fixes the units for $\VQQ$ and $r$.
}
\label{plot:V08}
\end{figure}

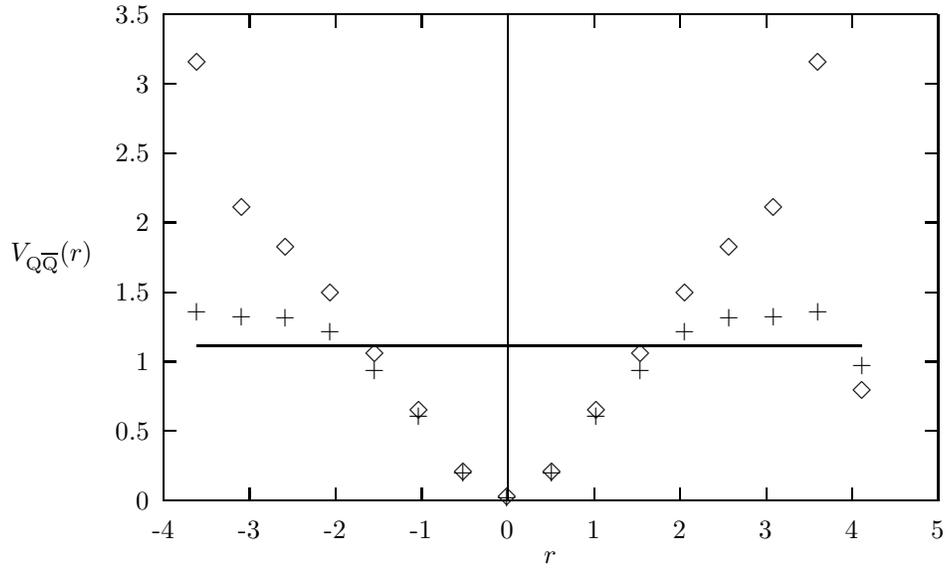
\begin{figure}
\setlength{\unitlength}{0.240900pt}
\ifx\plotpoint\undefined\newsavebox{\plotpoint}\fi
\sbox{\plotpoint}{\rule[-0.200pt]{0.400pt}{0.400pt}}%
\begin{picture}(1500,900)(0,0)
\font\gnuplot=cmr10 at 10pt
\gnuplot
\sbox{\plotpoint}{\rule[-0.200pt]{0.400pt}{0.400pt}}%
\put(220.0,113.0){\rule[-0.200pt]{292.934pt}{0.400pt}}
\put(760.0,113.0){\rule[-0.200pt]{0.400pt}{184.048pt}}
\put(220.0,113.0){\rule[-0.200pt]{4.818pt}{0.400pt}}
\put(198,113){\makebox(0,0)[r]{0}}
\put(1416.0,113.0){\rule[-0.200pt]{4.818pt}{0.400pt}}
\put(220.0,222.0){\rule[-0.200pt]{4.818pt}{0.400pt}}
\put(198,222){\makebox(0,0)[r]{0.5}}
\put(1416.0,222.0){\rule[-0.200pt]{4.818pt}{0.400pt}}
\put(220.0,331.0){\rule[-0.200pt]{4.818pt}{0.400pt}}
\put(198,331){\makebox(0,0)[r]{1}}
\put(1416.0,331.0){\rule[-0.200pt]{4.818pt}{0.400pt}}
\put(220.0,440.0){\rule[-0.200pt]{4.818pt}{0.400pt}}
\put(198,440){\makebox(0,0)[r]{1.5}}
\put(1416.0,440.0){\rule[-0.200pt]{4.818pt}{0.400pt}}
\put(220.0,550.0){\rule[-0.200pt]{4.818pt}{0.400pt}}
\put(198,550){\makebox(0,0)[r]{2}}
\put(1416.0,550.0){\rule[-0.200pt]{4.818pt}{0.400pt}}
\put(220.0,659.0){\rule[-0.200pt]{4.818pt}{0.400pt}}
\put(198,659){\makebox(0,0)[r]{2.5}}
\put(1416.0,659.0){\rule[-0.200pt]{4.818pt}{0.400pt}}
\put(220.0,768.0){\rule[-0.200pt]{4.818pt}{0.400pt}}
\put(198,768){\makebox(0,0)[r]{3}}
\put(1416.0,768.0){\rule[-0.200pt]{4.818pt}{0.400pt}}
\put(220.0,877.0){\rule[-0.200pt]{4.818pt}{0.400pt}}
\put(198,877){\makebox(0,0)[r]{3.5}}
\put(1416.0,877.0){\rule[-0.200pt]{4.818pt}{0.400pt}}
\put(220.0,113.0){\rule[-0.200pt]{0.400pt}{4.818pt}}
\put(220,68){\makebox(0,0){-4}}
\put(220.0,857.0){\rule[-0.200pt]{0.400pt}{4.818pt}}
\put(355.0,113.0){\rule[-0.200pt]{0.400pt}{4.818pt}}
\put(355,68){\makebox(0,0){-3}}
\put(355.0,857.0){\rule[-0.200pt]{0.400pt}{4.818pt}}
\put(490.0,113.0){\rule[-0.200pt]{0.400pt}{4.818pt}}
\put(490,68){\makebox(0,0){-2}}
\put(490.0,857.0){\rule[-0.200pt]{0.400pt}{4.818pt}}
\put(625.0,113.0){\rule[-0.200pt]{0.400pt}{4.818pt}}
\put(625,68){\makebox(0,0){-1}}
\put(625.0,857.0){\rule[-0.200pt]{0.400pt}{4.818pt}}
\put(760.0,113.0){\rule[-0.200pt]{0.400pt}{4.818pt}}
\put(760,68){\makebox(0,0){0}}
\put(760.0,857.0){\rule[-0.200pt]{0.400pt}{4.818pt}}
\put(896.0,113.0){\rule[-0.200pt]{0.400pt}{4.818pt}}
\put(896,68){\makebox(0,0){1}}
\put(896.0,857.0){\rule[-0.200pt]{0.400pt}{4.818pt}}
\put(1031.0,113.0){\rule[-0.200pt]{0.400pt}{4.818pt}}
\put(1031,68){\makebox(0,0){2}}
\put(1031.0,857.0){\rule[-0.200pt]{0.400pt}{4.818pt}}
\put(1166.0,113.0){\rule[-0.200pt]{0.400pt}{4.818pt}}
\put(1166,68){\makebox(0,0){3}}
\put(1166.0,857.0){\rule[-0.200pt]{0.400pt}{4.818pt}}
\put(1301.0,113.0){\rule[-0.200pt]{0.400pt}{4.818pt}}
\put(1301,68){\makebox(0,0){4}}
\put(1301.0,857.0){\rule[-0.200pt]{0.400pt}{4.818pt}}
\put(1436.0,113.0){\rule[-0.200pt]{0.400pt}{4.818pt}}
\put(1436,68){\makebox(0,0){5}}
\put(1436.0,857.0){\rule[-0.200pt]{0.400pt}{4.818pt}}
\put(220.0,113.0){\rule[-0.200pt]{292.934pt}{0.400pt}}
\put(1436.0,113.0){\rule[-0.200pt]{0.400pt}{184.048pt}}
\put(220.0,877.0){\rule[-0.200pt]{292.934pt}{0.400pt}}
\put(45,495){\makebox(0,0){$V_{\rm Q\overline{Q}}(r)$}}
\put(828,23){\makebox(0,0){$r$}}
\put(220.0,113.0){\rule[-0.200pt]{0.400pt}{184.048pt}}
\put(273,801){\raisebox{-.8pt}{\makebox(0,0){$\Diamond$}}}
\put(343,573){\raisebox{-.8pt}{\makebox(0,0){$\Diamond$}}}
\put(412,511){\raisebox{-.8pt}{\makebox(0,0){$\Diamond$}}}
\put(482,439){\raisebox{-.8pt}{\makebox(0,0){$\Diamond$}}}
\put(552,343){\raisebox{-.8pt}{\makebox(0,0){$\Diamond$}}}
\put(621,255){\raisebox{-.8pt}{\makebox(0,0){$\Diamond$}}}
\put(691,158){\raisebox{-.8pt}{\makebox(0,0){$\Diamond$}}}
\put(760,118){\raisebox{-.8pt}{\makebox(0,0){$\Diamond$}}}
\put(830,158){\raisebox{-.8pt}{\makebox(0,0){$\Diamond$}}}
\put(900,255){\raisebox{-.8pt}{\makebox(0,0){$\Diamond$}}}
\put(969,343){\raisebox{-.8pt}{\makebox(0,0){$\Diamond$}}}
\put(1039,439){\raisebox{-.8pt}{\makebox(0,0){$\Diamond$}}}
\put(1108,511){\raisebox{-.8pt}{\makebox(0,0){$\Diamond$}}}
\put(1178,573){\raisebox{-.8pt}{\makebox(0,0){$\Diamond$}}}
\put(1248,801){\raisebox{-.8pt}{\makebox(0,0){$\Diamond$}}}
\put(1317,285){\raisebox{-.8pt}{\makebox(0,0){$\Diamond$}}}
\put(273,410){\makebox(0,0){$+$}}
\put(343,402){\makebox(0,0){$+$}}
\put(412,400){\makebox(0,0){$+$}}
\put(482,378){\makebox(0,0){$+$}}
\put(552,317){\makebox(0,0){$+$}}
\put(621,246){\makebox(0,0){$+$}}
\put(691,156){\makebox(0,0){$+$}}
\put(760,117){\makebox(0,0){$+$}}
\put(830,156){\makebox(0,0){$+$}}
\put(900,246){\makebox(0,0){$+$}}
\put(969,317){\makebox(0,0){$+$}}
\put(1039,378){\makebox(0,0){$+$}}
\put(1108,400){\makebox(0,0){$+$}}
\put(1178,402){\makebox(0,0){$+$}}
\put(1248,410){\makebox(0,0){$+$}}
\put(1317,325){\makebox(0,0){$+$}}
\sbox{\plotpoint}{\rule[-0.400pt]{0.800pt}{0.800pt}}%
\put(273,356){\usebox{\plotpoint}}
\put(273.0,356.0){\rule[-0.400pt]{251.500pt}{0.800pt}}
\end{picture}
\caption{
 The potential defined in \eq{eq:VQQ} at $\gN/\mQ = .247$ for two heavy 
 quarks ($\Diamond$), and after the inclusion of one light flavor with
 $\mq/\mQ = .001$ ($+$).  The discrete momentum $K=15$ for both cases.  
 The horizontal line gives the additional energy $2\MQq - 2\mQ$ required 
 to produce two heavy-light mesons, each computed at $K=8$.  The square
 root of the string tension $\gNsq/2$ fixes the units for $\VQQ$ and $r$.
}
\label{plot:V12}
\end{figure}

The computer code produces masses and LC momentum-space wavefunctions
on whose Fourier transforms \eq{eq:schr} holds in the large-$\mQ$ limit.
Therefore, computing $\psi(r)$ 
and $\partial_r^2\psi(r)$ by transforming $\psi(q)$ and $q^2\psi(q)$ 
respectively and defining~\cite{lepage} 
\beq
\label{eq:VQQ}
\VQQ(r) \equiv {2\mu E \psi(r) + \partial_r^2\psi(r) \over \mu\gNsq \psi(r)} 
\eeq
should reproduce a linear potential in that limit.  
In particular, the degree to which $\VQQ$ conforms to $|r|$ indicates how 
accurately a nonrelativistic Schr\"odinger equation with a static potential 
describes this meson.  Figures \ref{plot:V08} and \ref{plot:V12} display 
potentials extracted in this manner for two values of $\gN/\mQ$.  These 
reconstruct $\gNsq |r|/2$ fairly reliably at the relatively modest momentum 
$K = 15$, although the effects of discretization appear in the turnover
of $\VQQ$ near its edges.  

\section{String Breaking}

In one spatial dimension, there are no transverse directions in
which the electric field may spread.  As a result, the Coulomb potential
is linear; in $\gauge$ gauge, it is all that remains of 
the gauge field.  Because the string is present in the classical Hamiltonian,
all the work on the quantum level goes into breaking it, making it a convenient
laboratory to study this effect.  To do so, I introduce an additional light-quark 
flavor and study the effect the presence of extra light-quark pairs has on the 
potential defined in \eq{eq:VQQ}.  More specifically, because the program includes 
all states consistent with quantum numbers specified, I include three flavors: 
two different heavy quarks with identical mass and one light quark, but I fix 
their respective flavor numbers to $1$, $-1$ and $0$.  This ensures that every 
state contains at least one heavy $\QQbar$ pair.

Figures \ref{plot:V08} and \ref{plot:V12} display preliminary results for
the potential in the presence of light quarks at couplings of 
$\gN/\mQ = .164$ and $.247$ and a light quark mass of $\mq = .001/\mQ$.  
The total discrete momentum $K=15$ for these,
leading to 15 points in the $\QQbar$ wavefunction.  I used an invariant mass
cutoff to exclude irrelevant extra heavy-quark pairs, and also restricted
by hand the state space to a single additional $\qqbar$ pair in order to
keep the number of states manageable; typically, about 1300.  Testing this
truncation by allowing additional pairs at lower $K$ had little effect, 
suggesting that the production of a single light-quark pair was predominantly
responsible for breaking the string.  The data displayed came from a run of 
roughly half an hour on a DEC alpha 3000/700 workstation.  Along with the ground 
states from which these potentials were extracted, the program produced the 
entire set of roughly 1300 masses and wavefunctions for this system.  The LC
Hamiltonian has a relatively simple structure, with the coupling and masses 
appearing as overall factors before the free and interacting terms.  Almost all of 
the computational effort goes into producing a complete orthonormal set of states 
and evaluating the Hamiltonian matrix.  Therefore, by storing 
separately the free and interaction matrices, reproducing the spectrum for 
different couplings and masses is essentially free.

Both the linear potential for the pure $\QQbar$ system and its breaking in 
the presence of light quarks are clearly evident.  The horizontal line in each 
plot indicates the additional cost in energy to produce a pair of heavy-light 
mesons, showing that the string breaks about where expected.  To obtain these, 
I computed the mass of the lightest $\Qqbar$ mesons with the same parameters but 
half the value of $K$.  This approximates that carried by each heavy-light 
meson produced in the $\QQbar\qqbar$ system.  

\begin{figure}
\setlength{\unitlength}{0.240900pt}
\ifx\plotpoint\undefined\newsavebox{\plotpoint}\fi
\begin{picture}(1500,900)(0,0)
\font\gnuplot=cmr10 at 10pt
\gnuplot
\sbox{\plotpoint}{\rule[-0.200pt]{0.400pt}{0.400pt}}%
\put(220.0,495.0){\rule[-0.200pt]{292.934pt}{0.400pt}}
\put(760.0,113.0){\rule[-0.200pt]{0.400pt}{184.048pt}}
\put(220.0,113.0){\rule[-0.200pt]{4.818pt}{0.400pt}}
\put(198,113){\makebox(0,0)[r]{-0.015}}
\put(1416.0,113.0){\rule[-0.200pt]{4.818pt}{0.400pt}}
\put(220.0,240.0){\rule[-0.200pt]{4.818pt}{0.400pt}}
\put(198,240){\makebox(0,0)[r]{-0.01}}
\put(1416.0,240.0){\rule[-0.200pt]{4.818pt}{0.400pt}}
\put(220.0,368.0){\rule[-0.200pt]{4.818pt}{0.400pt}}
\put(198,368){\makebox(0,0)[r]{-0.005}}
\put(1416.0,368.0){\rule[-0.200pt]{4.818pt}{0.400pt}}
\put(220.0,495.0){\rule[-0.200pt]{4.818pt}{0.400pt}}
\put(198,495){\makebox(0,0)[r]{0}}
\put(1416.0,495.0){\rule[-0.200pt]{4.818pt}{0.400pt}}
\put(220.0,622.0){\rule[-0.200pt]{4.818pt}{0.400pt}}
\put(198,622){\makebox(0,0)[r]{0.005}}
\put(1416.0,622.0){\rule[-0.200pt]{4.818pt}{0.400pt}}
\put(220.0,750.0){\rule[-0.200pt]{4.818pt}{0.400pt}}
\put(198,750){\makebox(0,0)[r]{0.01}}
\put(1416.0,750.0){\rule[-0.200pt]{4.818pt}{0.400pt}}
\put(220.0,877.0){\rule[-0.200pt]{4.818pt}{0.400pt}}
\put(198,877){\makebox(0,0)[r]{0.015}}
\put(1416.0,877.0){\rule[-0.200pt]{4.818pt}{0.400pt}}
\put(220.0,113.0){\rule[-0.200pt]{0.400pt}{4.818pt}}
\put(220,68){\makebox(0,0){-4}}
\put(220.0,857.0){\rule[-0.200pt]{0.400pt}{4.818pt}}
\put(355.0,113.0){\rule[-0.200pt]{0.400pt}{4.818pt}}
\put(355,68){\makebox(0,0){-3}}
\put(355.0,857.0){\rule[-0.200pt]{0.400pt}{4.818pt}}
\put(490.0,113.0){\rule[-0.200pt]{0.400pt}{4.818pt}}
\put(490,68){\makebox(0,0){-2}}
\put(490.0,857.0){\rule[-0.200pt]{0.400pt}{4.818pt}}
\put(625.0,113.0){\rule[-0.200pt]{0.400pt}{4.818pt}}
\put(625,68){\makebox(0,0){-1}}
\put(625.0,857.0){\rule[-0.200pt]{0.400pt}{4.818pt}}
\put(760.0,113.0){\rule[-0.200pt]{0.400pt}{4.818pt}}
\put(760,68){\makebox(0,0){0}}
\put(760.0,857.0){\rule[-0.200pt]{0.400pt}{4.818pt}}
\put(896.0,113.0){\rule[-0.200pt]{0.400pt}{4.818pt}}
\put(896,68){\makebox(0,0){1}}
\put(896.0,857.0){\rule[-0.200pt]{0.400pt}{4.818pt}}
\put(1031.0,113.0){\rule[-0.200pt]{0.400pt}{4.818pt}}
\put(1031,68){\makebox(0,0){2}}
\put(1031.0,857.0){\rule[-0.200pt]{0.400pt}{4.818pt}}
\put(1166.0,113.0){\rule[-0.200pt]{0.400pt}{4.818pt}}
\put(1166,68){\makebox(0,0){3}}
\put(1166.0,857.0){\rule[-0.200pt]{0.400pt}{4.818pt}}
\put(1301.0,113.0){\rule[-0.200pt]{0.400pt}{4.818pt}}
\put(1301,68){\makebox(0,0){4}}
\put(1301.0,857.0){\rule[-0.200pt]{0.400pt}{4.818pt}}
\put(1436.0,113.0){\rule[-0.200pt]{0.400pt}{4.818pt}}
\put(1436,68){\makebox(0,0){5}}
\put(1436.0,857.0){\rule[-0.200pt]{0.400pt}{4.818pt}}
\put(220.0,113.0){\rule[-0.200pt]{292.934pt}{0.400pt}}
\put(1436.0,113.0){\rule[-0.200pt]{0.400pt}{184.048pt}}
\put(220.0,877.0){\rule[-0.200pt]{292.934pt}{0.400pt}}
\put(45,495){\makebox(0,0){$\Delta\psi(r)$}}
\put(828,23){\makebox(0,0){$r$}}
\put(220.0,113.0){\rule[-0.200pt]{0.400pt}{184.048pt}}
\put(273,469){\raisebox{-.8pt}{\makebox(0,0){$\Diamond$}}}
\put(343,435){\raisebox{-.8pt}{\makebox(0,0){$\Diamond$}}}
\put(412,360){\raisebox{-.8pt}{\makebox(0,0){$\Diamond$}}}
\put(482,245){\raisebox{-.8pt}{\makebox(0,0){$\Diamond$}}}
\put(552,166){\raisebox{-.8pt}{\makebox(0,0){$\Diamond$}}}
\put(621,284){\raisebox{-.8pt}{\makebox(0,0){$\Diamond$}}}
\put(691,633){\raisebox{-.8pt}{\makebox(0,0){$\Diamond$}}}
\put(760,848){\raisebox{-.8pt}{\makebox(0,0){$\Diamond$}}}
\put(830,633){\raisebox{-.8pt}{\makebox(0,0){$\Diamond$}}}
\put(900,284){\raisebox{-.8pt}{\makebox(0,0){$\Diamond$}}}
\put(969,166){\raisebox{-.8pt}{\makebox(0,0){$\Diamond$}}}
\put(1039,245){\raisebox{-.8pt}{\makebox(0,0){$\Diamond$}}}
\put(1108,360){\raisebox{-.8pt}{\makebox(0,0){$\Diamond$}}}
\put(1178,435){\raisebox{-.8pt}{\makebox(0,0){$\Diamond$}}}
\put(1248,469){\raisebox{-.8pt}{\makebox(0,0){$\Diamond$}}}
\put(1317,478){\raisebox{-.8pt}{\makebox(0,0){$\Diamond$}}}
\end{picture}
\caption{
 The difference between the position-space $\QQbar$ wavefunctions $\psi(r)$
 computed at $\gN/\mQ = .247$ for two heavy quarks and after the 
 inclusion of one light flavor with $\mq/\mQ = .001$.  The square
 root of the string tension $\gNsq/2$ fixes the units for $r$.
}
\label{plot:WfnXdiff12}
\end{figure}
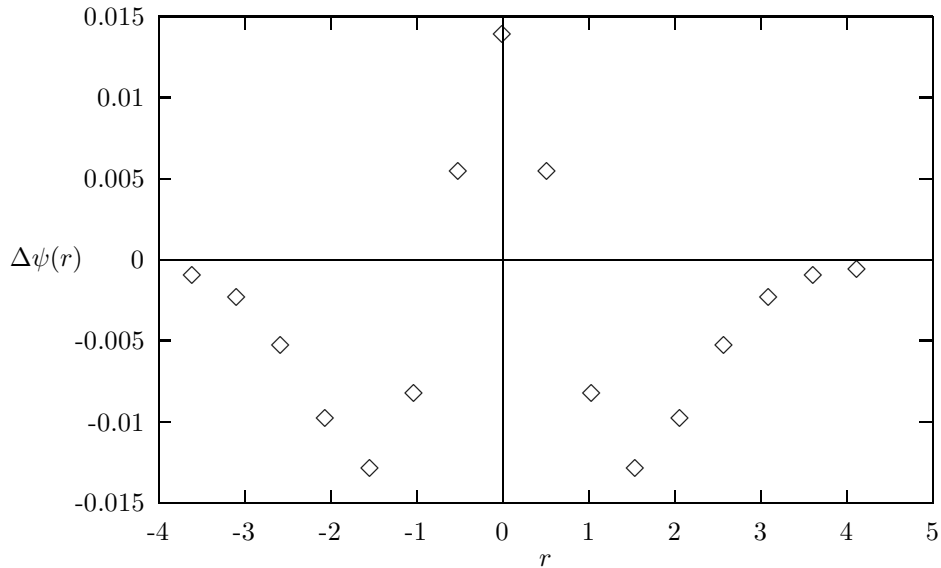

The ground state includes in its 
wavefunction not only the $\ket{\QQbar}$ states from which $\VQQ$ is extracted,
but also higher-Fock states $\ket{\QQbar\qqbar}$ with additional light
quark pairs.  By examining their magnitude, it is possible to gain some 
insight into which states are most responsible for breaking the string.  
For the $\gN/\mQ = .247$ case, states in which each light quark forms 
a singlet with a heavy quark and the LC momentum is carried
almost entirely and equally by the heavy quarks are predominant.
This is in accord with the tendency for Fock-state constituents to move 
with equal velocity \cite{ic}.  These account for approximately .5\% of the 
total probability.  
Figure \ref{plot:WfnXdiff12} shows the difference between the position-space 
wavefunctions for $\gN/\mQ = .247$, before and after the inclusion of 
light quarks.  
These exhibit the spread away from $r=0$ which results from charge 
screening by the light quarks.

In these figures, I computed $\VQQ$ from the ground state wavefunctions.  
Note, however, that \eq{eq:VQQ} applies equally to excited states. 
$\VQQ$ extracted from these is consistent but significantly degraded.

There are some tradeoffs evident in these preliminary results.  The
larger the coupling, the more dramatic the breaking and longer
the extent of the wavefunction.  But the larger coupling leads to a
less relativistic $\QQbar$ system, and a wider wavefunction in
momentum, making it more susceptible to finite box size errors.  It also
increases the importance of additional $\qqbar$ pairs thus far neglected.  

\section{Future Work}

In addition to simply increasing $K$, there are a number of ways of 
improving the accuracy of this study.  Fitting the wavefunctions to physically 
motivated continuous functions would ameliorate some of the discretization and 
finite-size errors.  The range of $\VQQ$, which is limited
by this method to the width of the wavefunction, could also be extended 
with the use of static sources, though this would mean abandoning the
momentum conservation currently built into the code.  Finally, an intelligent 
method for importance sampling of states, and perhaps the use of effective 
interactions to replace states excluded based on energy, would certainly
improve the efficiency of this program.

\section{Acknowledgements}

I would like to thank TJNAF and in particular G.\ Strobel for 
his generous hospitality in hosting this workshop in honor
of Stanley Brodsky's 60th birthday.  I cannot imagine having had 
a better advisor or collaborator than Stan, and am grateful to be 
able to make this contribution.  This work was supported by 
a grant from the Department of Energy.

\end{document}